\documentclass{emulateapj}
\usepackage{epsfig,natbib}
\usepackage{color}
\def\lap{\lower.5ex\hbox{$\; \buildrel < \over \sim \;$}}
\def\gap{\lower.5ex\hbox{$\; \buildrel > \over \sim \;$}}

\def\ergcm2s{${\rm erg\ cm^{-2}\ s^{-1}}$}

\def\ergscm2s{${\rm erg\ cm^{-2}\  s^{-1}}$}

\def\cm-2{${\rm cm^{-2}}$}

\begin{document}

\title{The Panchromatic Hubble Andromeda Treasury II. Tracing the
  Inner M31 Halo with Blue Horizontal Branch Stars}

\author{Benjamin F. Williams\altaffilmark{1},
Julianne J. Dalcanton\altaffilmark{1},
Eric F. Bell\altaffilmark{2},
Karoline M. Gilbert\altaffilmark{1,3},
Puragra Guhathakurta\altaffilmark{4},
Tod R. Lauer\altaffilmark{5},
Anil C. Seth\altaffilmark{6},
Jason S. Kalirai\altaffilmark{7},
Philip Rosenfield\altaffilmark{1},
Leo Girardi\altaffilmark{8}, 
}
\altaffiltext{1}{Department of Astronomy, Box 351580, University of Washington, Seattle, WA 98195; ben@astro.washington.edu; jd@astro.washington.edu; kgilbert@astro.washington.edu;philrose@astro.washington.edu}
\altaffiltext{2}{Department of Astronomy, University of Michigan, 550
Church St., Ann Arbor MI 48109; ericbell@umich.edu}
\altaffiltext{3}{Hubble Fellow}
\altaffiltext{4}{UC Santa Cruz; raja@uco.lick.org}
\altaffiltext{5}{NOAO; lauer@noao.edu}
\altaffiltext{6}{University of Utah; aseth@astro.utah.edu}
\altaffiltext{7}{Space Telescope Science Institute; jkalirai@stsci.edu}
\altaffiltext{8}{Padova, Italy; lgirardi@pd.astro.it}

\keywords{galaxies: individual (M31) --- galaxies: stellar populations --- galaxies: evolution}

\begin{abstract}

We attempt to constrain the shape of M31's inner stellar halo by
tracing the surface density of blue horizontal branch (BHB) stars at
galactocentric distances ranging from 2 kpc to 35 kpc.  Our
measurements make use of resolved stellar photometry from a section of
the Panchromatic Hubble Andromeda Treasury (PHAT) survey, supplemented
by several archival Hubble Space Telescope observations.  We find that
the ratio of BHB to red giant stars is relatively constant outside of
10~kpc, suggesting that the BHB is as reliable a tracer of the halo
population as the red giant branch. In the inner halo, we do not
expect BHB stars to be produced by the high metallicity bulge and
disk, making BHB stars a good candidate to be a reliable tracer of the
stellar halo to much smaller galactocentric distances.  If we assume a
power-law profile $r^{-\alpha}$ for the 2-D projected surface density
BHB distribution, we obtain a high-quality fit with a 2-D power-law
index of $\alpha=2.6^{+0.3}_{-0.2}$ outside of 3 kpc, which flattens
to $\alpha <$1.2 inside of 3 kpc.  This slope is consistent with
previous measurements but is anchored to a radial baseline that
extends much farther inward.  Finally, assuming azimuthal symmetry and
a constant mass-to-light ratio, the best-fitting profile yields a
total halo stellar mass of
2.1$^{+1.7}_{-0.4}\times10^9$~M$_{\odot}$.  These properties are
comparable with both simulations of stellar halo formation formed by
satellite disruption alone, and with simulations that include some
{\it in situ} formation of halo stars.

\end{abstract}

\section{Introduction}\label{intro}

The diffuse envelope of stars that appears to surround most or all
Milky Way-mass galaxies (massive galaxies, hereafter) encodes a wealth
of information about how such galaxies assembled and developed
\citep[e.g.][]{eggen1962,searle1978,bullock2001,bullock2005,zolotov2009,cooper2010}.
A major fraction of the stellar mass in such envelopes is expected to
be debris from tidally disrupted dwarf galaxies embedded in the dark
matter halos.  These disrupted galaxies are progressively incorporated
into the larger potential well of the main galaxy during its
hierarchical, bottom-up assembly. Accordingly, widespread effort has
been put into studying massive galaxy halos both observationally and
theoretically.

Over the past few decades, major observational progress has been made
in our understanding of the build-up of halos through discovery of
many dwarf galaxies around the Milky Way
\citep[e.g.,][]{ibata1994,willman2005,belokurov2007}, low surface
brightness structures in the Milky Way stellar halo
\citep{ibata1995,yanny2000,majewski2003,bell2008},and around the
outskirts of many nearby galaxies
\citep[e.g.,][]{malin1999,martinez-delgado2008,bailin2011,radburn-smith2011},
including M31
\citep[e.g.,][]{ibata2001,gilbert2007,mcconnachie2009,gilbert2009streams}.
The amount and character of such substructure at $r\ga 10$~kpc is both
qualitatively and quantitatively consistent with
cosmologically-motivated simulations in which stellar halos are built
up through the accretion of dwarf galaxies alone
\citep{bullock2005,bell2008,martinez-delgado2010,xue2011}.

At radii less than $\sim$20~kpc, it is expected that some fraction of
stellar halo mass may originate in the main potential well (often
called {\it in situ}), either formed early in the formation of the
galaxy or kicked up from the disk at later times through tidal
interactions
\citep[e.g.][]{kazantzidis2008,zolotov2009,font2011,brook2012}.
Studies attempting to disentangle the fraction formed in situ 1-2 are
challenging both observationally and theoretically for a variety of
reasons.

Observationally, the stellar halo has both low surface brightness and
relatively low stellar mass, and so is easily overwhelmed by higher
surface brightness components of a galaxy, making accurate
measurements challenging.  A further (related) difficulty is that the
separation of a halo from a bulge component (where the bulge in great
part is expected to have formed the bulk of its stars in the main
galaxy potential well) has been extremely difficult because disk and
bulge dominate the stellar halo \citep[][hereafter
C11]{dorman2012,courteau2011}, leading a number of groups to consider
both components together
\citep[e.g.][]{irwin2005,kaliraihalo2006,sarajedini2012,font2011}.

A related theoretical problem is that in numerical simulations
relatively rare modes of star formation at relatively high
temperatures and low densities, which are not yet well-characterized,
may contribute disproportionately to low-mass diffuse stellar halos
\citep[see discussions in][]{zolotov2009,munshi2012}. The correct star
formation prescription for the halo gas is still uncertain, making the
simulated stellar halo properties less reliable.  In what follows, we
attempt to give a fresh perspective to this issue by focusing on the
possibility of using blue horizontal branch (BHB) stars in M31 as a
probe of M31's inner stellar halo.

\subsection{Current M31 Stellar Halo Measurements} 

Fits to the surface brightness profile of M31 do not constrain the
shape of the inner halo.  The surface brightness profile is dominated
by the red giant branch (RGB) stars in M31.  These studies fit all
three galaxy components (disk, halo, and bulge) simultaneously, but
must assume a halo model inside of $\sim$10~kpc where all 3 components
contribute significantly.  Not only are the fits to the inner halo
profile model dependent, the fits by C11 show clear degeneracies
between the different components, making extrapolating the halo
profile inwards from 10 kpc unreliable.  Thus, the modeling of the
inner halo can result in significant variations of the stellar halo
mass fraction at these inner radii.

Kinematic decompositions of the halo outside of 10~kpc, performed with
spectra of individual RGB stars, currently suggest a 2-D projected
surface density power-law index of ${\sim}{-}2$ \citep[K. Gilbert et
al. ApJ, submitted; ][hereafter G05]{raja2005}.  However, M31's bulge
and halo are both dynamically hot and exhibit metallicity and/or age
gradients \citep[e.g.][]{kaliraihalo2006,gilbert2007},
making it extremely difficult to disentangle M31's bulge from M31's
halo closer to the galaxy center, even with kinematics of individual
red giant branch stars.

Furthermore, a case can be made that it is dangerous to extrapolate
the outer halo stellar profiles inwards to small galactocentric
distances in M31, or in any other galaxy. Inside of $\sim$35~kpc, the
stellar populations of the M31 halo are substantially more metal-rich
than those of the Milky Way and have a range of ages
\citep{durrell1994,durrell2001,durrell2004,brown2006,brown2007,brown2008,richardson2008}.
In addition, there are deviations from a single power law in the
surface brightness profile of the outer stellar envelope of M31
\citep{kaliraihalo2006}.  

Taken together, these differences have led some workers to suggest
that at radii $\la 35$~kpc one should think of the extended stellar
envelope as an extended bulge instead of halo, or one should simply
lump both components together into a spheroid. In addition,
simulations of stellar halo formation, both those that are composed
only of tidal debris from dwarf disruptions and those that include an
{\it in situ} component, show changes in power law profiles with
radius, frequently flattening in profile towards the smallest radii
\citep[e.g.][]{zolotov2009,cooper2010,font2011}. Without
reliable constraints on the shape of the inner halo, comparisons of
the observed halo to simulations or estimates the stellar halo mass
are hampered.

In short, there are currently no straight-forward tools for mapping
the halo profile in the inner galaxy. Thus any current knowledge about
stellar halo profiles is limited to large radii. 

\subsection{Using BHB Stars to Trace the Halo}

Our approach to recovering the inner structure of the halo is to
identify a tracer unique to the stellar population of the halo. More
specifically, we attempt to define the stellar halo as the population
that is responsible for the presence of field blue horizontal branch
(BHB) stars.  We then check the consistency between our results and
previous halo studies.  We attempt this new approach in order to avoid
the difficulties of more direct kinematic or photometry measurements.

The halo measurements of previous studies have been limited to the
outskirts of the galaxy to ensure they are measuring a halo population
\citep[e.g.][]{brown2008,richardson2008,kaliraihalo2006}.  In these
regions, previous works have found a total stellar population that is
poorly fit with a single power-law profile, motivating a description
of the outer parts of M31 in terms of a stellar halo (often taken as
synonymous with a power law component), an extended bulge (a Sersic
index component prominent out to large radius), and an extended disk
\citep[e.g.][C11]{ibata2005,kaliraihalo2006}.

In contrast to current surface brightness and spectroscopic
techniques, the BHB may provide a way to trace the halo profile to
small radii, breaking degeneracies between the disk, bulge, and halo
galaxy components. The BHB is the part of the horizontal branch just
blueward of the RR Lyrae instability strip. It is made of low-mass
stars in the phase of central helium burning, with effective
temperatures from $\sim$7200--$\sim$40000~K. The BHB is routinely
subdivided by temperature into three parts, the HBA
($\sim$7200~K$\lesssim$T$_{\rm eff}{\lesssim}$12000~K), HBB
(12000~K$\lesssim$T$_{\rm eff}{\lesssim}$20000~K), and EHB (T$_{\rm
eff}{\gtrsim}$20000~K) \citep[for a full review on HB evolution
see][and references therein]{Catelan2009}. In this work we include
only the HBA, which is the section typically associated with
metal-poor populations, when referring to the BHB. For a typical old
and metal-poor halo environment, these BHB stars will spend their
$\sim$100 Myr lifetime confined within a horizontal strip just 0.2 mag
wide in M$_{\rm bol}$, so that the BHB location can be approximated by
a simple zero-age horizontal branch (ZAHB) sequence, as the one
illustrated in Figure~\ref{cmds}.

Although BHB stars are typically associated with low metallicity
populations, it is also possible to make hot HB stars at high
metallicities.  Enhanced mass loss, and possibly high helium content,
generate a second formation channel; however this metal-rich channel
is more likely to produce extremely hot HB stars, not normal HBA
stars, as witnessed by the metal-rich open cluster NGC6791
\citep[e.g.][]{kalirai2007}, and as demonstrated by the very low
formation efficiency of RR~Lyrae among metal-rich populations
\citep{layden1995}.  BHB stars instead, are likely to have
metallicities similar to the RR~Lyrae, i.e. with a mean
[Fe/H]${\sim}{-}1$ \citep{pietrukowicz2012} and just a very minor tail
of stars extending to high metallicities (see also Figure~\ref{gcs}).
For this reason, these stars have been successfully used as probes of
the Galactic halo stellar population \citep{preston1991}.  As a
result, a prominent BHB component is unlikely to be formed by the
higher metallicity disk and bulge populations.  We note the
possibility of a contribution from the thick disk, which has recently
been measured to have a metallicity similar to that of the halo
\citep{collins2011}, but argue against this possibility on the basis
of previous M31 disk population studies and kinematic arguments
\citep[see \S~\ref{disk}, also][who find a very low fraction of local
BHB stars originating in the thick disk]{kinman2009}.

There is clearly a BHB component in the M31 stellar halo
\citep{vandenbergh1991}, as easily seen in deep HST CMDs in the
literature \citep[e.g.][]{bellazzini2003,brown2003}.  Furthermore,
RR~Lyrae stars have been cataloged in M31 far from the central bulge
\citep[e.g.,][]{bernard2012}.  We note that the BHB stars we are using
to probe the halo are distinct from extreme horizontal branch (EHB)
stars, which are much hotter that the traditional BHB.  These EHB
stars are common in metal-rich populations including the M31
\citep[e.g.][]{rosenfield2012}. and Galactic bulges
\citep[e.g.][]{oconnell1999,busso2005} and therefore would not be
appropriate tracers of the halo.  Furthermore, EHB stars are quite
faint in the optical, as their spectral energy distribution peaks in
the far-UV, making them fall below the completeness limit in our ACS
data.

Herein we use BHB stars to trace the M31 halo component to small
galactocentric distances, where the overall fraction of halo stars is
low and difficult to measure any other way. We isolate the BHB in the
M31 field population and measure its surface density to galactocentric
distances as small as 1.6 kpc from the center of M31. Our measurement
of the halo profile provides the means to break degeneracies with
other components, as well as to estimate the total halo stellar mass.
In \S~2 we describe our data set and measurement techniques.  In \S~3
we present the resulting BHB density profile and argue that it is
reasonable to assume that they are all halo members.  We then quantify
our constraints on the halo profile and compare these to
currently-available decomposition measurements and simulations.
Finally, in \S~5 we provide a brief summary of our results and their
place in our understanding of the structure of M31.  

We assume a distance of 780~kpc \citep{stanek1998} for all conversions
from angular distances to kiloparsecs.  All power-law indexes refer to
2-D projected surface density profiles.  We make no corrections for
inclination, as there is no clear evidence in the literature that the
M31 stellar halo is non-spherical, we have no way of knowing if an
inclination correction is needed.  If the halo is indeed significantly
flattened and inclined, it would affect the magnitude ranges adopted
for our BHB selection in the outer halo fields by $\sim$0.1 mag.  Such
an offset would not significantly impact our surface density
measurements; however, flattening would affect our mass estimate, as
discussed in \S~\ref{mass}.

\section{Data Acquisition and Analysis}

\subsection{Panchromatic Hubble Andromeda Treasury Data}

As part of the PHAT survey \citep{dalcanton2012}, we obtained HST/ACS
data in F475W and F814W covering a large fraction of the M31 disk.
All of the PHAT data for this paper were acquired and analyzed as part
of the survey, as detailed in \citet{dalcanton2012}.  In short,
photometry was performed using the package DOLPHOT
\citep{dolphin2000}.  The output was filtered to reject non-point
sources and low-quality measurements, as detailed in
\citet{dalcanton2012}.  Artificial star tests were performed on each
field to allow measurement of completeness as a function of color and
magnitude.

In studying the survey photometry, we noticed a blue feature extending
across the vertical main-sequence in a few of our survey fields (see
Figure~\ref{cmds}).  Further examination of the areas exhibiting this
feature revealed that they all fell within a small corner of the
survey, shown in Figure~\ref{footprints}.  This corner lies at the
farthest point that the survey reaches along the M31 minor axis, and
thus has the highest ratio of halo to disk stars.

\subsubsection{Counting BHB Stars}

\begin{figure}
\centerline{\epsfig{file=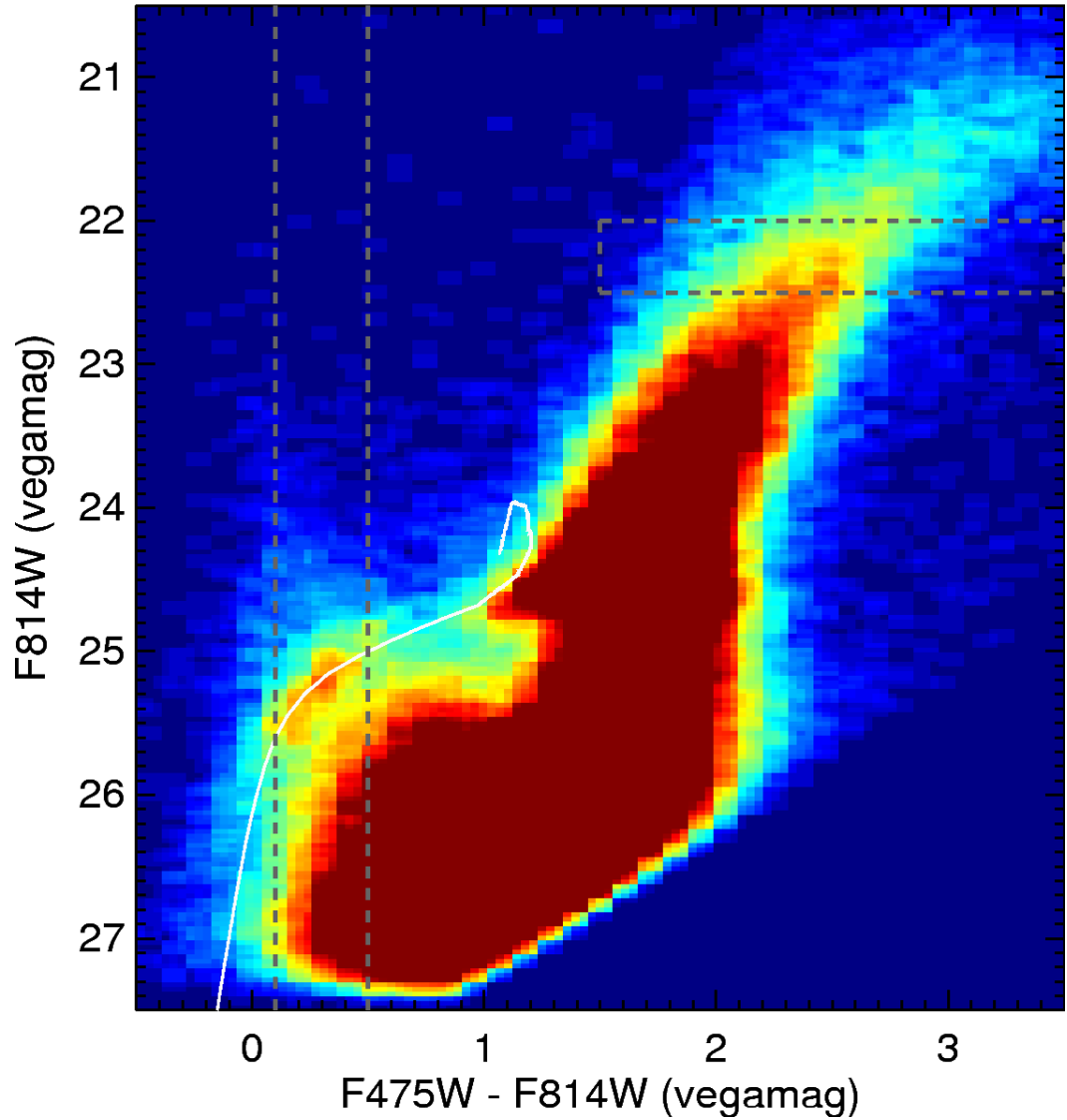,width=5.5in,angle=0}}
\caption{The color-magnitude diagram of the first PHAT field where we
noticed the BHB ($R_{minor}$=3.8~kpc).  Redder colors denote higher
densities of data points.  Overplotted in white is the theoretical
zero age horizontal branch for [Fe/H]= -1.7 from the Padova stellar
evolution models. The overdensity in the CMD associated with the BHB
is clearly evident as the density peak coincident with the model at
F475W$-$F814W$\sim$0.3, F814W$\sim$25.3.  The metallicity has little
effect in this sequence at the color of the BHB, but changes the shape
of the red end.  Overplotted in dashed gray lines are the boundaries
over which we fit histograms in order to measure the properties of the
BHB stars. The dashed gray box outlines the area in which we counted RGB
stars for BHB/RGB ratios.}
\label{cmds}
\end{figure}

\begin{figure}
\centerline{\epsfig{file=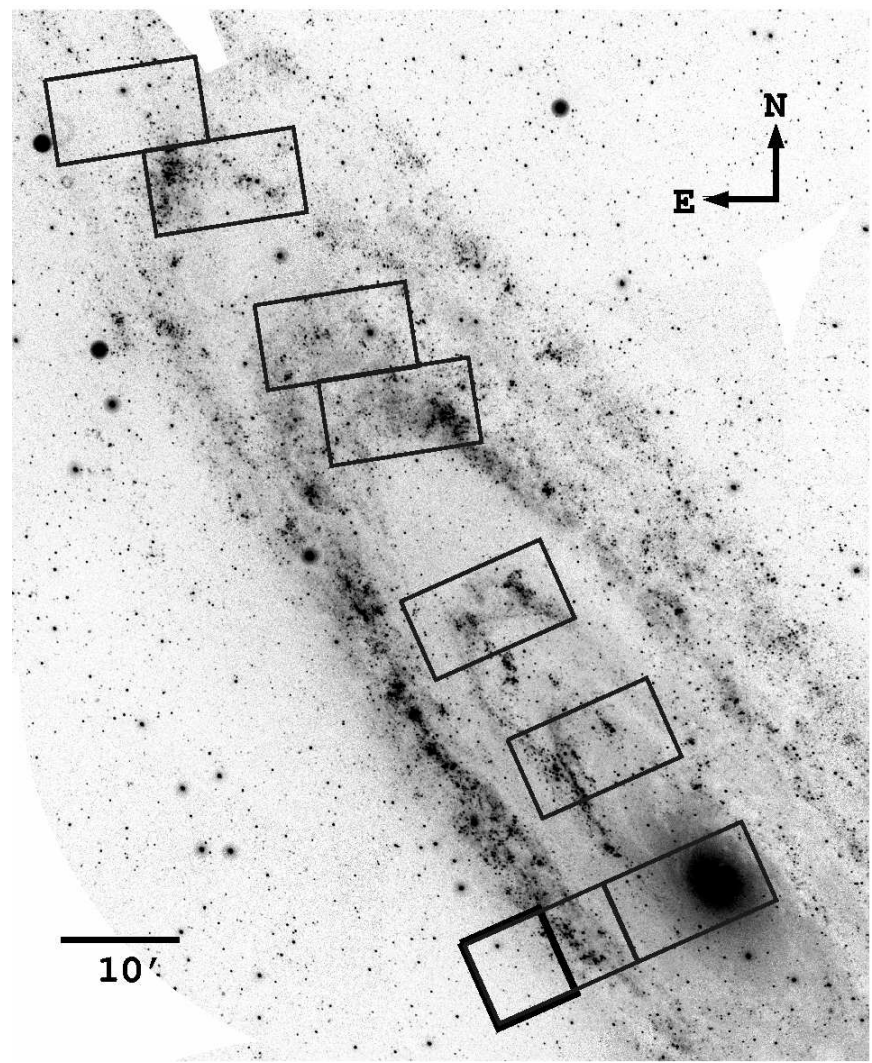,width=4.5in,angle=0}}
\caption{The northern half of the GALEX NUV image of M31.  Thin line
boxes show the regions of the PHAT survey completed when the BHB was
seen.  Thick black box shows the region where the BHB was noticed in the
photometry: the area farthest out on the minor axis, with the lowest
disk contribution.}
\label{footprints}
\end{figure}

\begin{figure}
\centerline{\epsfig{file=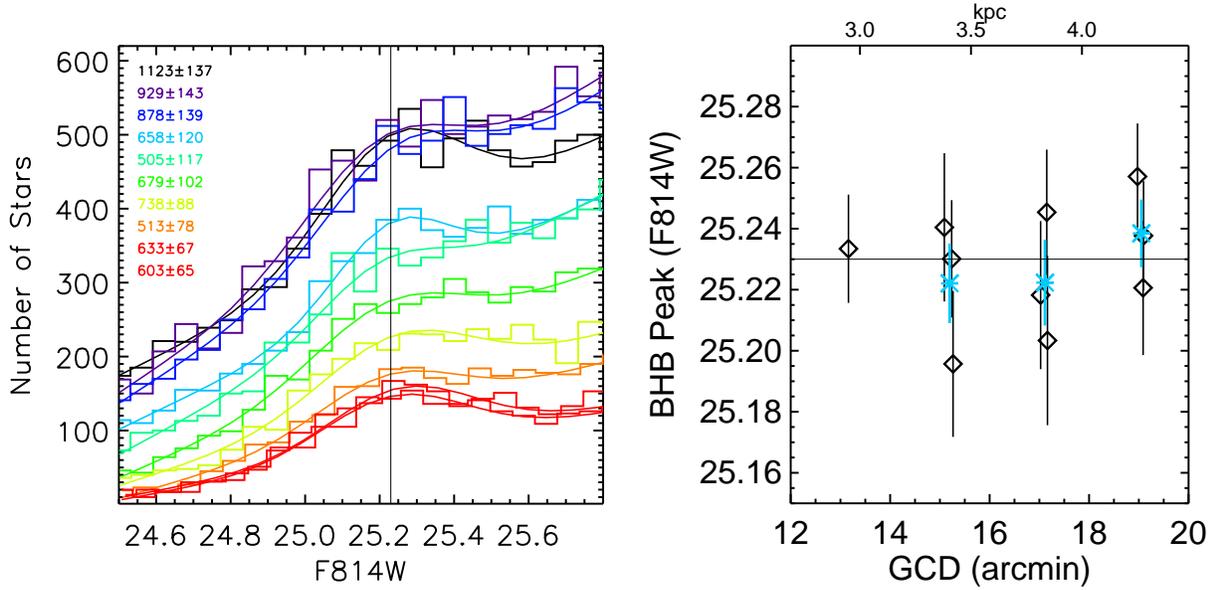,width=6.5in,angle=0}}
\caption{{\it Left:} Luminosity functions for stars with
  0.1$<$F475W$-$F814W$<$0.5, in PHAT fields with clear BHB features.
  Overplotted are the best-fitting line+Gaussian functions.  Each
  field is represented by a different color.  The numbers in the upper
  left indicate the number of stars (and uncertainty) in the Gaussian
  component. {\it Right:} The resulting best-fit apparent F814W
  magnitude for the center of the BHB as a function of galactocentric
  distance.  Cyan points show the combined measurements of multiple
  fields at similar radii.  There is no trend in BHB peak magnitude
  with radius or surface brightness in M31, thus the Gaussian fits are
  not significantly affected by crowding or completeness, and that the
  detected stars in each location are behind equal amounts of
  extinction. The BHB in M31 has F814W=25.23.}
\label{gaussfits}
\end{figure}

We isolated the BHB component by taking a vertical cut in the CMD at
0.1$<$F475W$-$F814W$<$0.5 (see dashed vertical lines in
Figure~\ref{cmds}), and plotting F475W and F814W magnitude
  histograms. Assuming $E_{B-V}$=0.1, this intrinsic color range
  (0.1$<$F475W$-$F814W$<$0.5 after correcting for reddening)
  corresponds to 7200K--9600K.  Our major axis disk field luminosity
  functions using this color selection was well-fit with a simple
  straight line, suggesting that the underlying main sequence
  population has a simple linear luminosity function.  We therefore
  fit these luminosity functions, which clearly contained a
  significant enhancement at the magnitude expected for the BHB, with
  a straight line plus a Gaussian; the latter component was then used
  as a measure of the magnitude and strength of the BHB feature at
  this color. We chose our color range based on the data in
  Figure~\ref{cmds}.  At bluer colors, the BHB sequence extends
  vertically in our filter set, blending completely with the main
  sequence. Redward of our BHB selection window, the RR~Lyrae
  instability strip spreads the feature, also making the BHB difficult
  to isolate.  In F814W, we found the peak magnitude to be consistent
  across all fields where we measured it; however, in F475W, the peak
  magnitude was slightly less consistent, owing to small amounts of
  dust reddening.  Therefore we show the less dust-affected F814W fits
  in Figure~\ref{gaussfits}.  Finally, we attempted rotated LFs that
  were orthogonal to the BHB in this color range.  We found in these
  cases the background LF became more complicated as did comparisons
  across different filter sets.  In the end, the simple F814W
  luminosity function provided the most reliable measurements across
  our full sample.

The line component of our fits represents the upper main-sequence
stars that also occupy this color-magnitude region.  The number of BHB
stars was obtained by subtracting the line component away from the
total, thus removing the upper main-sequence contamination from our
BHB sample.

We performed checks to ensure that our sample was not strongly
affected by the presence of dust in the M31 disk.  First, we checked
that the apparent magnitude of the BHB was consistent across all
fields.  While there was some increased scatter in F475W, it was small
($<$0.1 mag), and the scatter was even smaller in F814W ($<$0.03
mag). In neither case was a trend with galactocentric distance seen.
Furthermore, we inspected the PHAT infrared photometry of all of the
regions where we detected the BHB, and found the red giant branch to
have a single, narrow tip in all regions.  We also checked the
reddening value for the nearest globular cluster to the region, which
is very low \citep[B201; E$_{(B-V)}$=0.04$\pm$0.02;][]{fan2008}.
Finally, we inspected the {\it Spitzer} 24$\mu$ maps of the regions,
finding no significant emission above the background.  Therefore, with
all of these tracers, there is no evidence for the presence of a
significant dust layer at these radii along the minor axis that would
affect our measurements of the BHB.

\subsubsection{BHB Upper Limits}

Outside of 2.5~kpc on the minor axis and 5~kpc on the major axis,
artificial star tests show that our 50\% completeness limit is a
magnitude or more below the BHB feature, making it simple to isolate
and measure the BHB component.  In contrast, we were not able to
isolate the feature inside of 2.5~kpc on the minor axis and 5~kpc on
the major axis, because our photometric uncertainties and completeness
affected our data too strongly at the magnitudes of interest.
Attempts to solve this problem by applying completeness corrections
from false star tests were not effective, due to the sharp cutoff in
completeness in our region of interest.  Essentially, when
completeness corrections become greater than a factor of 2, the
inferred number of stars has associated uncertainties that our fitting
routine could not overcome.  Therefore, at these inner radii, we
assume that the BHB is located in the same color-magnitude range as in
all of the fields where it was measured directly (25.1$<$F814W$<$25.3
and 0.1$<$F475W$-$F814W$<$0.5).  We then counted the total number of
stars in this range and corrected for the mean completeness.  This
calculation gives us the total number of stars which potentially could
be BHB stars.  However, because we cannot reliably subtract off the
contribution from the MS or post-AGB stars, we quote our measurements
in these fields as upper limits.

Our interpretation of the inner radius measurements as upper limits is
unaffected by magnitude biases from crowding.  In very crowded fields,
such as the inner M31 bulge, our photometry shows biasing towards
brighter magnitudes.  Such biasing would put the BHB stars brightward
of the magnitude range within which we are counting BHB stars.
Because the luminosity function increases toward fainter magnitudes,
measuring the completeness-corrected numbers of stars at a brighter
magnitude interval (to take such biasing into account) would only
serve to {\it lower} our upper limits, making them less conservative.
In one case, our completeness-corrected LF showed a small peak at
F814W=25.05, slightly brightward of our assumed BHB magnitude.  We
verified that shifting our window 0.1 mag brightward
(25.0$<$F814W$<$25.2) to include this small peak did not significantly
impact our results.  Such a change only slightly shifted our
upper-limits, and the lowest upper limit changed by $<$3\%.  Thus our
upper limits are conservative and are not strongly sensitive to the
precise magnitude cuts.

We attempted to augment our radial baseline by applying the same
measurement techniques to other fields. We first attempted to detect
the BHB in the PHAT survey data along the major axis.  However, we
were unable to isolate a BHB feature because the expected number of
BHB stars from the halo is small compared to the number of
main-sequence stars present in the disk component along the major
axis.  These fields had such high contamination levels that
upper-limits measured using our blind counting technique were so large
that they proved to be of no use.  We therefore do not report any
measurements taken near the major axis.

\subsection{Relevant Archival Data}

In order to greatly increase our radial baseline, we searched for
relevant archival data that would contain resolved photometry of the
BHB in the M31 halo.  First, we found the deep photometry of the halo
released as high level science products \citep[PIDs: 9453, 10265,
10816, PI: Brown,][hereafter B08]{brown2009,brown2008}.  Then we found
deep archival imaging away from the major axis or known streams that
we could quickly process with the PHAT pipeline (PID: 10394, PI:
Tanvir, \citealp{tanvir2012}; PID: 11362, PI: Rich)

\subsubsection{Public Deep Halo Photometry}

We downloaded the very deep photometry catalogs of B08, from ACS
fields at 11, 21, and 35 kpc out along the minor axis, as well as one
located at R=20 kpc in a known stream.  These fields are well away
from the disk, contain stars at the location of the BHB, and have
no main-sequence stars..

For these data, which use a different filter set and different
photometric system, we found that the feature was isolated enough that
a comparable measure of the number of BHB stars (leaving out the
instability strip and the vertical extension) was obtained from
counting all stars with
$-1.0{<}$F606W$_{STMAG}{-}$F814W$_{STMAG}{<}{-}$0.6 and
26.0$<$F814W$_{STMAG}{<}$26.7.  The difference in F814W magnitude
corresponds simply to the difference in F814W zeropoint between
VEGAMAG (25.5) and STMAG (26.8).  The color range was chosen by
looking at the CMDs to find the color corresponding to the BHB
feature; however, it is similar to the expected shift of ${\sim}{-}1$ (a
formal transformation of F475W$-$F814W=0.3 results in
F606W$_{STMAG}{-}$F814W$_{STMAG}{=}-0.7$).

 Although some of these fields are known to lie in streams, the
fraction of stream stars is well-constrained by kinematics
\citep{gilbert2009}.  For the 11~kpc field and 20~kpc stream field,
only 56$\pm$16\% and 25$\pm$10\% of the stars belong to the
kinematically hot halo \citep{gilbert2009}. These fractions are
determined by spectroscopy of RGB stars in fields that overlap the HST
imaging.  The kinematics of stars in these fields display distinct
cold streams with low dispersion as well as a population of stars with
high dispersion (hot) halo component; maximum-likelihood fits to the
line-of-sight velocity distributions in each field yield estimates of
the fractions of stars in the hot and cold components.  While both of
these fields are in known streams, the outer field lies on a more
overdense stream, resulting in a lower fraction of hot halo stars.
Therefore, we scale our measurements in these fields by 0.56 and 0.25
respectively, to account for the overdensities, and we include the
associated errors in our uncertainties.  We note that accounting for
these overdensities is necessary because of the small sample area of
the ACS field of view at the distance of M31, as well as the bias of
HST programs to point at streams.  Larger areas in random halo
locations, as one would obtain in more distant systems, would include
both overdense and underdense regions, averaging out to the true mean
density of the halo.  In addition, there is some chance that the
streams and the rest of the halo have different BHB properties in M31,
as has been seen in the Galaxy \citep[e.g.][]{bell2010}; however, the
similar F$_{BHB}$ values seen through the streams and halo (see
Figure~\ref{radialfits}) suggests that, if such differences are
present in M31, they do not significantly affect our study.

\subsubsection{Other Relevant Archival Observations}

We searched for other archival ACS observations suitable for our
project.  Although there have been many ACS observations in the M31
halo, most were either located on known overdense streams or were too
close to the major axis to provide a clean measurement of the BHB.
Unlike the B08 fields, other archival stream fields did not have
spectroscopically determined stream fractions.  Therefore we only
included fields outside of known streams.

We found 2 fields with favorable locations and depth.  These were a
parallel observation in program 11632, which was looking at background
QSOs (we refer to this field as 11632\_M31-HALO-SE), and an
observation of a halo globular cluster (we refer to this field as
10394\_M31-HALO-NW).  For this latter field, a $20''{\times}20''$
region centered on the globular cluster we exclude from our analysis.

These 2 fields were taken in F606W and F814W like the B08 fields;
however, we reduced them in our own pipeline using a stricter quality
cut appropriate for less crowded fields. In crowded fields, the
crowding parameter is high for most of the stars because they all have
close neighbors.  In sparse fields, crowding is very sensitive to
background galaxies because shredded background galaxies have
photometry similar to crowded stars.  This contaminant is
insignificant in crowding-limited fields with hundreds of thousands of
real stars, but is significant in sparse fields with tens of thousands
of real stars.  Much cleaner CMD features (such as the BHB) are gained
by limiting the crowding cut in sparse fields.  We produce CMDs
similar in quality to those of B08 at the BHB by applying a crowding
cut of 0.1 to these halo fields.  Our artificial star tests showed
that even with this more conservative cut, we are complete at the
location of the BHB in these sparse fields.

\section{Results}

\subsection{BHB Magnitude}

Our measurements of the BHB feature are detailed in
Table~\ref{measurements}.  These include, for each ACS field, the median (projected) galactocentric
distance of the stars from the center of M31, the
apparent magnitude of the BHB peak in our color range,
the number of BHB stars, the number of RGB stars in our selection
region, and the resulting BHB/RGB fraction.  The measurements are plotted in
Figures~\ref{gaussfits}--\ref{radialfits}.  We find that the feature
has an apparent magnitude of F814W=25.23$\pm$0.03, at a F475W$-$F814W
color of 0.3.  Assuming a distance modulus of 24.47 \citep{stanek1998}
and foreground extinction of $A_V{=}0.21$ \citep{schlegel1998}, this
corresponds to M$_{F814W}$=0.63$\pm$0.05.
The magnitude and color are consistent with the model BHB shown in
Figure~\ref{cmds}, which has $[Fe/H]{=}-1.7$; however, all model
BHBs with $-2.3{<}[Fe/H]{<}{-}1.0$ are similar to our observed BHB.

\subsection{BHB/RGB Fraction}

There is also information contained in the relative number of stars in
the BHB feature compared to more well-populated features.  For
example, since BHB stars are typically produced by low metallicity
populations, there may be a relation between the fraction of BHB stars
present in an old population and the population's metallicity.  To
assess the relative strength of the BHB feature in our data, we
calculated the ratio of the number of stars in the Gaussian component
of the histogram fit (N$_{BHB}$) to the number of stars with
22.0$<$F814W$<$22.5 and 1.5$<$F475W$-$F814W$<$3.5 (N$_{RGB}$).  We
took the RGB sample from 23.8$<$F814W$_{STMAG}\,<$23.3 in the B08
data.  This magnitude slice provides a good proxy to the total stellar
mass in the field, as it is dominated by the very well populated RGB.
We tailored the magnitude range to yield a fraction
($F_{BHB}\,\equiv\,N_{BHB}/N_{RGB}$) close to unity for $\sim$12~Gyr
old clusters with [Fe/H]${\sim}{-}1$. These values allow for easy
comparisons between the halo light fraction and $F_{BHB}$ (see
\S~\ref{bhbfrac})

Once we had defined F$_{BHB}$ to measure the strength of the BHB in
our data, we checked its sensitivity to population metallicity by
measuring it for a large sample of well-studied globular
clusters.  This comparison was performed to look for similarity
between the M31 halo and Galactic globular clusters of similar age and
metallicity.  Since the Galactic halo, and most of the globular
clusters, significantly differ in age and metallicity from the M31
halo, only the higher metallicity and somewhat younger end of the
globular clusters is directly comparable.  However, we include the
full  ACS globular cluster treasury program
\citep{sarajedini2007,dotter2010} for context.  

Specifically, we measured the BHB fractions of the Galactic globular
clusters (GCs) of the ACS globular cluster treasury program
\citep{sarajedini2007,dotter2010} All GC CMDs were corrected for their
different distances and extinctions using the values from
\citet{dotter2010}.  Values for metallicity and age for NGC~6388 were
taken from \citet{worley2010} and for NGC~6441 from
\citet{gratton2006}. Distances and reddening values for these two
clusters were taken from \citet{harris1996}.  We then took our BHB
from the CMD region with $-$0.1${<}M_{V}{-}M_{I}{<}0.3$ and
$-0.3{<}M_{I}{<}0.8$, and our RGB from the CMD region with
$-2.0{<}M_{I}{<}{-}1.5$.  The results are shown in Figure~\ref{gcs}.
Applying the mean metallicities and ages from the full-field CMD
analysis of B08, we find that the BHB fraction of the M31 halo is in
good agreement with the F$_{BHB}$-[Fe/H] correlation for Galactic GCs,
despite the broad metallicity spread contained in the M31
fields.\footnote{Although the B08 metallicities are not from
spectroscopy, they are in agreement with spectroscopic metallicity
measurements of stars in the vicinity.}  F$_{BHB}$ values for the M31
halo ($\sim$0.7) are similar to those measured for GCs of similar ages
(11 Gyr) and metallicities ($-$1.0 -- $-$0.5).  Furthermore, we note
that F$_{BHB}$ is the same in fields inside and outside of streams,
suggesting that the streams have similar F$_{BHB}$ to the
kinematically hot halo.  This finding is consistent with the
full-field analyses of B08, which found the ages and metallicities of
the steam field to be similar to other halo fields.

Figure~\ref{gcs} confirms that at old ($\gap$10~Gyr) ages strong BHB
populations are associated with low metallicities.  Below
[Fe/H]${<}{-}$1.5, we find F$_{BHB}$ is universally high
(1$<$F$_{BHB}{<}$6) with a median of 3.2.  At higher metallicities,
the strength of the BHB population drops dramatically, falling by a
factor of 10 as the metallicity rises to [Fe/H]${<}{-}$0.5. Above this
metallicity, the BHB is unlikely to exist at all.  Finally, we note
that this comparison includes the moderately metal-rich globular
clusters NGC~6388, NGC~6441, and Lynga~7, which exhibit blue
horizontal branches, showing that these clusters do not significantly
affect the overall behavior of F$_{BHB}$ with age and metallicity.

We can use the data in Figure~\ref{gcs} to place limits on the
metallicity distribution of the M31 stellar halo.  If we assume that
all of the BHB stars come from a metal-poor tail in the population and
that the outer fields from the archive are pure M31 halo, we can infer
from Figure~\ref{gcs} that no more than $\sim$10\% of the halo
population is likely to belong to a metal-poor than with
[Fe/H]${<}-$1.5.  Such a tail would produce F$_{BHB}$ values higher by
a factor of 2 in these pure halo fields. This result is also
consistent with the full CMD analyses summarized in B08, which have
little stellar mass at [Fe/H]${<}{-}$1.5. We note that this result
relies on the assumption that F$_{BHB}$ behaves in a similar way in
M31 and the Galaxy.  Figure~\ref{gcs} is consistent with such an
assumption; however, there have been studies that suggest otherwise
\citep[e.g.][]{rich2005}.

Finally, Figure~\ref{gcs} shows us that the youngest (and most
metal-rich) halo field appears to have a somewhat high F$_{BHB}$ value
for its metallicity, making it consistent with the values from the
other fields.  Thus, our measurements suggest that the F$_{BHB}$ value
in the halo is relatively independent of radius.  Therefore, if the
halo indeed has a gradient in its mean metallicity, as has been
suggested in several works \citep[e.g.][]{kaliraihalo2006,brown2008},
then the gradient does not appear to strongly affect F$_{BHB}$ in the
radial range considered here.  This result is consistent with the fact
that the BHB surface density measurements obey a single-sloped
power-law profile from 3--35~kpc in Figure~\ref{radialfits}.  

\subsection{BHB Surface Density Profile}

We now analyze the surface density profile of BHB stars.  If the
stellar halo can be defined as the population that is responsible for
the field BHB stars, then this profile should be consistent with
previous measurements of the outer halo.  We show our measurement of
the surface density profile of BHB stars in M31 in the left panel of
Figure~\ref{radialfits}.  The surface density falls off with
galactocentric distance following roughly the surface brightness
profile of the halo measured by G05 and C11 outside of 11~kpc.
However, the BHB profile is steeper between 3~kpc and 11~kpc than
inward extrapolations of such current halo models. Interestingly, our
data suggest a single power-law slope all the way from 3~kpc to
35~kpc. The slope is steeper than (but within the uncertainties of)
G05, and it is similar in slope to the outer halo portion of the fits
of C11.  Thus, our data are providing a new constraint on the shape of
the inner halo, suggesting the single power-law slope extends inward
to 3~kpc before flattening.

We fit to the BHB surface density profile using the halo
parametrization of C11.  When fitting the data, we only applied our
lowest measured upper limit by setting the value and uncertainty both
equal to half of the upper limit.  The power-law function from C11
provides an excellent fit to the BHB data ($\chi^2_{\nu}$=0.97).

$$
\Sigma(r) = \Sigma_*\left[\frac{(1+R_*/a_h)^2}{(1+r/a_h)^2}\right]^{{\alpha}/2}
$$

\noindent
Where $\Sigma(r)$ is the BHB surface density at radius $r$, $\alpha$
is the projected 2D spatial density distribution power-law index, and
$a_h$ is a core radius (in units of kpc) inside of which the profile
flattens. $\Sigma_*$ and $R_*$ are normalization constants. Not
surprisingly, the best fit has the surface density right at our lowest
upper limit.  Our data constrain the model to
$\alpha$=2.6$^{+0.3}_{-0.2}$ and $a_h$=2.7$^{+1.0}_{-0.8}$~kpc.
While this functional form features a core and a central slope that
goes to zero as $r\rightarrow0,$ we do not claim the actual detection
of a core in the inner halo, only that the slope decreases to $<$1.2
interior to 3 kpc ( a steeper profile would exceed our upper limit at
1.6~kpc). We only use this functional form to aid in comparison with
previous measurements.

How do our halo parameters compare with other measurements? We first
compare with other estimates of power law slopes.  C11's surface
brightness decomposition study gives a power law index of
${\alpha}$=2.52$\pm$0.08, in excellent agreement with our power
law. G05 finds ${\alpha}={\sim}-2$ with no quantitative uncertainties,
and thus roughly consistent with our result. A more formal result from
a spectroscopically confirmed sample of M31 halo stars yields
${\alpha}=2.2{\pm}0.2$ (K. Gilbert et al., ApJ, submitted), also
consistent with our measurement.  We have also estimated a flattening
radius for the BHB profile (a robust measurement), which as long as
the BHB/RGB ratio for the halo is constant (an assumption) is
consistent with a core radius for the stellar halo as a whole. Our
estimate does not agree with others in the literature (C11 have a
large core, attributing instead much of that light to the disk
component; and G05 lack a core at all in their parametrized profile),
but the constraints from other works were necessarily weak due to the
small fraction of halo stars at these small radii.

\subsection{The Mass of M31's Stellar Halo}\label{mass}

We can integrate our 2-D projected surface density profile to obtain
an estimate of the total halo stellar mass.  First, we make the strong
assumption that the BHB surface density is directly proportional to
the halo stellar mass surface density, effectively making $\Sigma(r)$
the profile of the stellar halo surface density. Taking $\mu_V$=29 mag
arcsec$^{-2}$ at 21 kpc (from integrating the 21~kpc field catalog of
B08), M$_{V_{\odot}}$=4.82 and assuming a stellar mass to V-band light
ratio of 1.5 \citep{bellazzini2012}, we calculate
$\Sigma_*$=1.4$\times$10$^5$~M$_{\odot}$~kpc$^{-2}$ and $R_*$=21~kpc.

Then, we assume azimuthal symmetry, making the total mass equal to the
following integral:

$$
M_{tot} = \int \Sigma(r)dA = \int^{2\pi}_0\int^{260}_0
r\Sigma(r)~drd\theta
$$

\noindent
This calculation yields a total halo stellar mass out to the virial
radius \citep[260~kpc][]{seigar2008} of
2.1$^{+1.7}_{-0.4}{\times}10^{9}$~M$_{\odot}$, with a half-mass radius
of 7~kpc.  Changing the outer boundary of the integration has a small
effect on the resulting value.  Integrating to 40~kpc (near the outer
boundary of our photometry data) yields
1.8${\times}10^{9}$~M$_{\odot}$, while integrating to infinity yields
2.3${\times}10^{9}$~M$_{\odot}$.  We also note that many other stellar
halos/envelopes appear to be oblate within $10-20$~kpc with axis
ratios $c/a$ between 0.4 and 0.7 (e.g., the Milky Way:
\citealp{juric2008}; NGC 253: \citealp{bailin2011}; M81:
\citealp{barker2009}, M. Vlajic et al. in prep.; NGC 2403:
\citealp{barker2012}), it is possible that our assumption of azimuthal
symmetry will turn out to be incorrect. In that case, we expect that
the stellar mass estimate would scale very approximately as $a/c$ (the
inverse of the projected axis ratio of the stellar halo). This simple
scaling applies to the idealized case, where the M31 minor axis is
aligned with that of the minor axis of the spheroid and there are no
projection effects.

\subsubsection{Comparing with Other Stellar Halo Measurements}

Comparing with previous estimates of M31's halo mass, our measurement
is slightly on the high side, but broadly consistent with
expectations. We note that our estimate is less than an inward
extrapolation of the power-law (index = 2.2$\pm$0.2) of Gilbert et
al. (ApJ, submitted), or other similar results
(e.g. index=2.17$\pm$0.15, \citealp{tanaka2010}; index=1.91$\pm$0.11,
\citealp{ibata2007}; index$\sim$-2.3, \citealp{irwin2005}) which
results in a halo stellar mass (assuming no core and excluding the
cusp inside of 1 parsec) of $\sim$1.3$\times$10$^{10}$~M$_{\odot}$ out
to the virial radius and a half-mass radius of $<$1~kpc.  If we apply
the $a_h$ value measured from the BHB profile to the Gilbert et
al. (ApJ, submitted) profile, the resulting halo stellar mass will be
the consistent with our value (1.8${\times}10^{9}$~M$_{\odot}$, but
with a half-mass radius of $\sim$16~kpc).  Assuming M/L$\sim$2, the
measurement of \citet{ibata2007} yields
$\sim$2${\times}10^{9}$~M$_{\odot}$.  \citet{irwin2005} estimate
$\sim$2.5\% of the total light from M31 is from the halo, and C11
estimates 4\%.  Assuming a total M31 luminosity of 2${\times}10^{10}$
L$_{B{\odot}}$ and M/L$_{\rm B}{\sim}2$, these correspond to stellar
halo masses of 1.0--1.6${\times}10^{9}$~M$_{\odot}$, respectively.

We can also compare our M31 stellar halo mass with those of other
nearby large disk galaxies.  Our estimate of the stellar mass of the
halo of M31 is less than current estimates for the mass of the stellar
halo of the nearby disk galaxy NGC~253
\citep[4${\times}10^{9}$~M$_{\odot}$][]{bailin2011}, but larger than
current estimates for the mass of the stellar halo of the Milky Way
out to 40~kpc (3.7${\pm}1.2{\times}10^{8}$~M$_{\odot}$,
\citealp{bell2008}; 2--10${\times}10^{8}$~M$_{\odot}$,
\citealp{siegel2002,juric2008,deason2011}).

\subsubsection{Comparing with Simulations}

M31 and the Milky Way live in massive dark matter halos of similar
mass \citep[1.4$\times$10$^{12}$~M$_{\odot}$][]{watkins2010}.  The
stellar mass of the Milky Way disk is estimated to be
5.5$\times$10$^{10}$~M$_{\odot}$ \citep{flynn2006}, and that of M31 is
estimated to be roughly the same when determined from the absolute
visual magnitude of M$_V{\sim}{-}21.0$ \citep{font2011}. Thus,
simulations of Milky Way-like galaxies can be compared to M31
observations as well as Milky Way observations.  

Comparing to cosmologically-motivated simulations in which the stellar halo
is built up through the accretion of dwarf galaxies alone, we find a
reasonable degree of agreement between the properties we infer for M31's
stellar halo and BJ05 and C10. BJ05 find halo stellar masses $\sim 2 \times
10^9 M_{\sun}$, with 3-D power law slopes outside 10kpc of between $-$2.5
and $-$3.5.  C10 find a somewhat wider range in possible halo masses,
between $10^8 M_{\sun}$ and $2.5 \times 10^{9} M_{\sun}$, and 3-D power law
slopes ranging between $-2$ and $-4.5$. Hydrodynamical models of stellar
halo formation that include an {\it in situ} component find rather higher
stellar masses, 0.5--3.4$\times$10$^{10}$~M$_{\odot}$
\citep{zolotov2009}\footnote{\citet{font2011} do not quote stellar halo
masses as all the properties quoted in their work are for bulge$+$halo.}.
Possibly relevant given the strong evidence of a very active M31 merger
history
\citep[e.g.][]{gilbert2009streams,mcconnachie2009} is that both
\citet{zolotov2009} and \citet{font2011} find that the galaxies with the
most active merger histories have the lowest {\it in situ} fractions in
their simulations. Yet, simulated stellar halos have large halo-to-halo
scatter which is merger history-dependent; given this scatter, we conclude
at this stage that the properties of M31's stellar halo appear to be
consistent with the range of stellar halos expected in a cosmological
context.

\subsection{BHB/RGB Ratio Profile}\label{bhbfrac}

\begin{figure}
\centerline{\epsfig{file=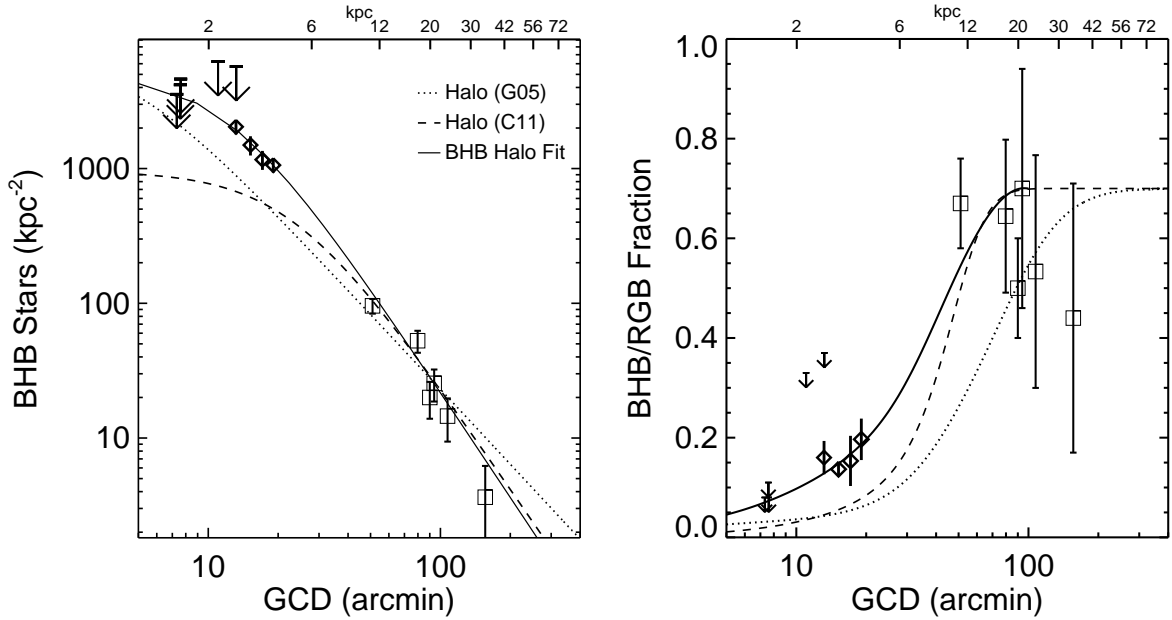,width=6.5in,angle=0}}
\caption{{\it Left: }The number density of BHB stars as a function of
M31 galactocentric distance.  Diamonds show combinations of the
multiple measurements at similar radii in the PHAT data.  Squares show
the B08 and archival photometry. Upper-limits are shown with downward
arrows. Dotted line: M31 halo profile from G05 normalized to match our
data at 20 kpc. Dashed line: M31 halo profile from C11 normalized to
match our data at 20 kpc.  Solid line: a fit of the BHB data using the
power-law halo parametrization of C11 ($\chi^2_{\nu}{=}0.97$).  {\it
Right:} F$_{BHB}$ as a function of galactocentric distance over the
same radial range. Lines are from the same models as {\it Left} but
represent 0.7$\times\,H/T$ where $H$ is the Halo light and T is the
total light in the two models. The low BHB/RGB ratios near the center
are due to the bulge and disk components contributing substantially to
the RGB but not to the BHB, and do not reflect metallicity changes in
the halo component. The factor of 0.7 scales the fraction to the pure
halo F$_{BHB}$ value (see \S~\ref{bhbfrac}). The curve corresponding
to the BHB fit (solid curve) assumes the total from \citet{raja2005}
to compute fractions and is only computed out to the ``pure halo''
value (0.7) at 21 kpc.}
\label{radialfits}
\end{figure}

\begin{figure}
\centerline{\epsfig{file=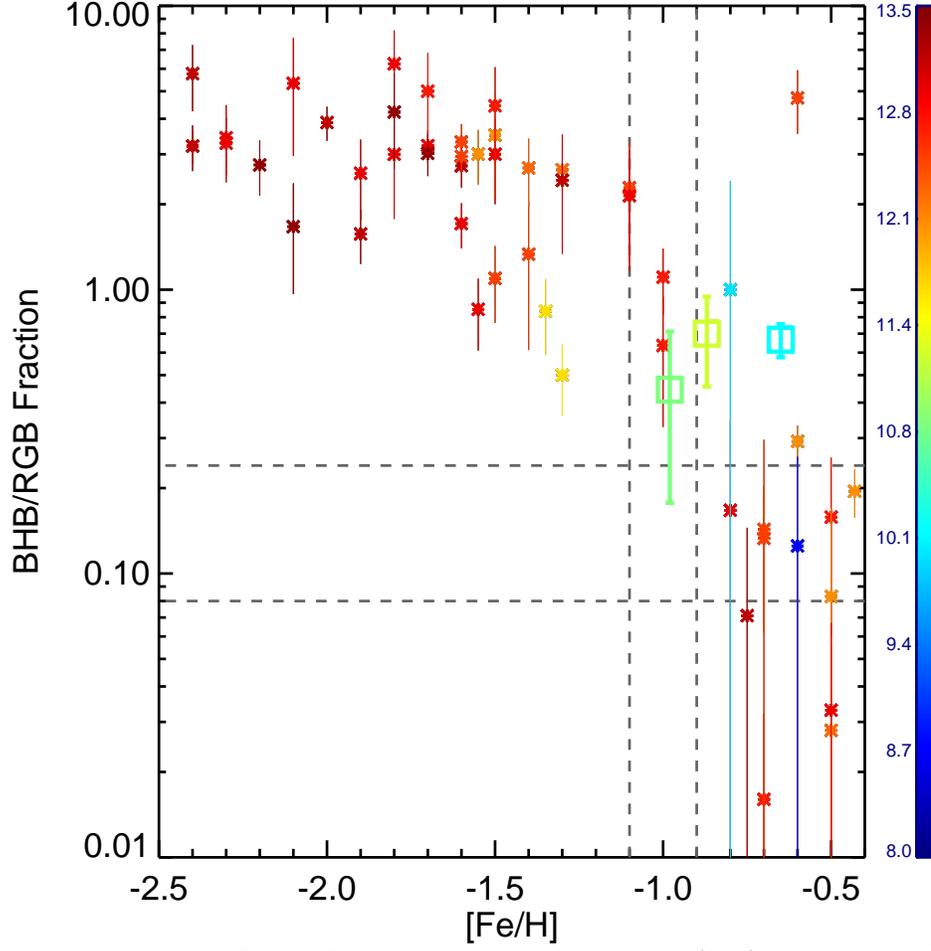,width=4.5in,angle=0}}
\caption{BHB fractions measured for Galactic GCs from the sample of
\citet{dotter2010} using the same CMD regions as for the M31 halo
fields of B08.  Points are color-coded by age, given by the color bar
in Gyr.  GCs are shown with asterisks.  M31 points are shown with open
squares using the ages and metallicities measured in B08.  Dashed
vertical lines mark the metallicity of the M31 thick disk
\citep{collins2011}.  Dashed horizontal lines lines mark the range
observed at 3--5 kpc from the galaxy center.  We assume that the low
values at these inner radii are in large part due to the disk and
bulge components adding to the RGB, but not to the BHB.}
\label{gcs}
\end{figure}

We can use the data in the right panel of Figure~\ref{radialfits} to
assess the fraction of halo stars at small radii.  In the right panel
of Figure~\ref{radialfits}, we plot F$_{BHB}$ as a function of
galactocentric distance, along with the G05 and C11 models of the
fraction of total light coming from the M31 halo.  The models suggest
that the relative strength of the halo clearly increases with
galactocentric distance out to 11~kpc, then remains relatively
constant at the pure halo value. This radial distribution suggests
that the population becomes dominated by halo members outside of
$\sim$11~kpc.

The behavior of F$_{BHB}$ in Figure~\ref{radialfits} is consistent
with what we expect for a halo population based on the G05 and C11
models.  At large radii, the measured values of F$_{BHB}$ are
consistent with being flat (although the uncertainties are large, due
to the small number of stars).  At smaller radii, however, the
strength of the BHB feature falls dramatically compared to the RGB.
Empirically, this drop indicates the increasing contribution of old
stellar populations that do not host BHB stars.  If F$_{BHB}$ for the
M31 halo is roughly constant with galactocentric distance, then the
low values inside of 11~kpc are due to the increased RGB contribution
from the bulge and disk components.  If we assume that the BHB stars
are all halo members and that F$_{BHB}$ for the halo is 0.7, then the
expected behavior of F$_{BHB}$ is to follow 0.7$H/T$ where $H$ is the
halo stellar surface density and T is total stellar surface density.

Perhaps not surprisingly, both the radial distribution of F$_{BHB}$
and the number density profile suggest that the BHB stars inside of
4.5~kpc do not fit inward extrapolations of current models of the halo
profile. Inside of 5~kpc, F$_{BHB}$ is significantly larger than
expected.  This excess in BHB stars suggests a number of
possibilities, which we have explored, or address in the subsections
that follow.  1) The halo has a different shape from inward
extrapolations of current decompositions (the interpretation we
explored in sections 3.3 and 3.4). 2) F$_{BHB}$ of the halo is not
constant.  3) The bulge is contaminating the BHB sample.  4) The disk
is contaminating the BHB sample.  We now discuss these latter 3
possibilities and argue that they are unlikely, leaving us to conclude
that we have indeed identified a means by which to constrain the shape
of the inner stellar halo.

\subsection{Does the Halo Produce BHB Stars at all Radii?} 

An assumption involved in our analysis is that the M31 stellar halo
produces field BHB stars in the same fraction at all radii.  While our
results from 10--35~kpc suggest that this assumption is reasonable,
there is no clear way to test the assumption at small radii.  Indeed,
our upper-limits inside of 2~kpc allow the possibility that the there
is no population that produces a field BHB at these small radii. We
stress that our profile only describes the population that produces a
field BHB, which we suggest {\it may be} a reasonable way to define
the M31 stellar halo.  Thus, in this analysis, all kinematically hot
stars beyond those required to produce the BHB profile shape
(i.e. many of the stars inside of 3~kpc) are associated with the M31
bulge.

Even if the halo produces Field BHB stars at all radii, F$_{BHB}$ of
the halo component may not be constant.  The outer M31 halo (outside
of $\sim$10~kpc) is known to have a metallicity gradient
\citep{kaliraihalo2006}.  However, this gradient is {\it negative}
(metallicity increases with decreasing radius).  While it is known
that the {\it total} HB fraction (including the RHB) increases with
increasing metallicity \citep{salaris2004}, based on the behavior of
F$_{BHB}$ in Figure~\ref{radialfits}, F$_{BHB}$ decreases with
increasing metallicity.  Thus, we would expect the {\it stellar halo
only} F$_{BHB}$ to decrease slightly toward the galaxy center.  Such
behavior would make F$_{BHB}$ {\it even lower} than the dashed or
dotted curves in Figure 4 near the center, which is the opposite of
the observed behavior.  Therefore, a changing F$_{BHB}$ of a halo of
the type currently in the literature appears an unlikely explanation
for the observed radial profile.  We note that if the halo $F_{BHB}$
does decrease towards smaller radii, the stellar mass and power law
slopes of the halo inferred in sections 3.3 and 3.4 would increase. We
now move on to discuss the possibility of bulge or disk contamination.

\subsection{Are the BHB Stars From the Stellar Halo?}

An important assumption involved in our use of the BHB as a probe of
the M31 stellar halo is that neither the bulge nor the disk components
contribute significantly to the field BHB.  Here we argue that our
data suggest this assumption is reasonable.  As detailed below, if we
assume that a significant fraction of BHB stars belongs to either the
bulge or the disk, we can fit the profile, but only if these
components contain old, metal-poor populations that have not been
observed any other way.  Furthermore, the density of BHB stars from
3--35~kpc appears to follow a single power-law distribution.  Thus,
the BHB stars appear to belong to the stellar halo, which extends to
small radii with a slightly different structure than inward
extrapolations of currently-available decomposition models.  Inside of
$\sim$3~kpc, simple inward extrapolations greatly overpredict the
observed numbers of BHB stars, suggesting a break in the profile.

\subsubsection{Possible Contributions of Bulge BHB stars}\label{bulge}

One possible source of excess BHB stars at small galactocentric
distances is the bulge. However, given the degree to which the bulge
dominates at small radii, the low numbers of BHB stars in M31 seen
inside of 2~kpc suggests that the bulge population is not responsible
for the excess.  Furthermore, the high-metallicity of the bulge
(0.0$\lap$[Fe/H]$\lap$0.4 out to 1~kpc, \citealp{saglia2010}) also
argues against it as the source of the BHB stars seen at 2.7--4.5~kpc.
On the other hand, the bulge is known to contain a broad range of
metallicities \citep{sarajedini2005}.  If we assume that some fraction
of the bulge population is metal-poor enough to produce a significant
BHB ([Fe/H]${<}{-}$0.5), then we can estimate what fraction of metal-poor
stars would best reproduce the BHB profile.  If the resulting estimate
is reasonable, then our assumption that the BHB traces the halo
becomes questionable.

The potential contribution of the bulge to the BHB is different
depending on the M31 bulge profile.  Therefore we tested both the
bulge profiles of C11 and G05, which bracket the possibilities of halo
size. The C11 bulge is small, and the G05 bulge is large.

First, assuming the decomposition models of C11, we can calculate the
bulge properties necessary to reproduce the observed BHB density
profile.  The preferred C11 fit has a small bulge component that
decreases very steeply to zero by a minor axis distance of 4.5~kpc.
Given this decomposition fit, it is impossible for the bulge to be
responsible for the BHB stars because there would not be enough
stellar mass associated with the bulge at 3--5 kpc to produce the
observed number of BHB stars, even if the bulge metallicity were as
low as the halo metallicity.

If instead we assume the decomposition model of G05 (large bulge), we
can produce a BHB profile consistent with the observations if the
bulge has a steep increase in the fraction of metal-poor stars outside
of 1~kpc. If we assume that some fraction of the bulge population is
metal-poor enough to produce a BHB ([Fe/H]${<}{-}$0.5), we can
estimate what fraction of metal-poor stars would best reproduce the
BHB profile. Our very conservative upper-limits at these radii
(1.6~kpc) suggest that less that 10\% of the inner bulge population is
metal-poor enough to produce a BHB.  We can bring the profiles in
complete agreement by imposing a fraction of metal-poor stars in the
M31 bulge that increases {\it very} steeply, from 0\% to 50\% from the
galaxy center to 2.5~kpc.  Another way to bring the profiles into
agreement would be to decrease the metallicity of the metal-poor
component of the bulge.  This explanation is possible, but it is not
likely, as it would be odd if the metal-poor stars of the bulge were
of lower metallicity than those of the halo.  Since no such steep
metallicity gradient has been observed at these radii in the bulge, it
appears unlikely that the BHB stars at 2.7--4.5~kpc are dominated by
the bulge component.

\subsubsection{Possible Contributions of Disk BHB stars}\label{disk}

Another possible source of excess BHB stars is the outer disk
component. However, there is a very powerful empirical argument
against disk stars being responsible for the prominent BHB population
at 2.7$-$4.5~kpc along the minor axis: the BHB is not detected in the
outer fields along the {\it major axis} of the PHAT survey data, as
would be expected if the disk stellar populations are well-mixed (as
an old population like the BHB is expected to be).  The deprojected
disk radius of 2.7--4.5~kpc is 1--1.5 degrees. The PHAT data only
extend to $\sim$1 degree along the major axis, but the B08 disk field
($\sim$2 degrees out on the major axis) shows very little evidence for
the old ($>$10~Gyr), metal-poor component \citep{brown2006} needed to
produce a BHB excess.  Therefore it appears that the disk is unlikely
to be a significant source of BHB stars. It is worth noting that this
lack of BHB stars is not necessarily because the metallicity of the
outer disk isn't low enough, but rather may indicate that the vast
majority of outer disk stars are too young to produce a prominent BHB.
Furthermore, kinematic measurements suggest a low disk contribution at
these minor axis radii \citep[$\lap$10\% at 9~kpc]{gilbert2007}.
Thus, both the observed populations of the disk itself and the
spectroscopic disk fraction argue against a disk origin of the BHB
stars at these radii.

\section{Conclusions}

We have measured the number density of BHB stars in M31 at
galactocentric distances ranging from 1.6~kpc to 35~kpc using
photometry from the PHAT survey along with archival halo ACS fields.
Our measurements show that the properties of the BHB of M31 are
consistent across two degrees of galactocentric distance.  Galactic
globular clusters with ages of $\sim$10~Gyr and metallicities of
$-1.0{\lesssim}$[Fe/H]${\lesssim}{-}0.5$ (similar to those of the M31 halo
values measured from deep CMD analysis) have BHB properties that match
the M31 halo as well.  Taken together, these measurements suggest the
BHB is a useful tool for tracing galaxy halos.   

In additional tests, we showed that the BHB surface density profile
follows that of the known halo outside of 10~kpc.  However, there is
an excess of BHB stars at 2.7--4.5~kpc over inward extrapolations of
current M31 halo profiles.  To match the radial profile of the BHB a
combination of halo and bulge and/or disk components can be applied;
however, the contributions of these components that is required is
inconsistent with the expected properties of the M31 bulge and disk
populations at these radii.  While it is possible that the disk and/or
bulge populations could be contaminating our BHB sample, there is no
compelling evidence that the M31 bulge or disk harbors a significant
BHB population at any radius.  Furthermore, these components would
need to be contributing BHB stars in just the right proportions for
the BHB density to follow a power-law density distribution.  Such a
conspiracy is not impossible, but seems unlikely.  Thus, our use of
the BHB to trace the inner halo is further justified.

The BHB data are well matched by a power-law with an index of
$2.6^{+0.3}_{-0.2}$ outside of 3 kpc decreasing to $<$1.2 inside of 3
kpc.  This profile describes the population that produces a field BHB,
which we suggest is a reasonable way to define the M31 stellar halo.
In this picture, all kinematically hot stars that do not follow this
profile could be associated with the M31 bulge. Our profile slope is
consistent with the range in slopes characteristic of simulations of
stellar halo formation in a cosmological context.

Normalizing our best-fitting profile function to a halo stellar mass
density of 1.4$\times$10$^5$~M$_{\odot}$~kpc$^{-2}$ at 21 kpc yields a
total stellar halo mass of
2.1$^{+1.7}_{-0.4}{\times}10^{9}$~M$_{\odot}$ for M31. This mass
compares well with, but is on the high side of, other estimates of the
M31 halo stellar mass, and it is significantly higher than current
estimates of the stellar halo mass of the Milky Way.

About a decade ago, the first large stream in M31 was discovered
\citep{ibata2001}.  Since then, the M31 halo has been mapped to large
radii with Keck (K. Gilbert et al., ApJ, submitted) and the CFHT
\citep{mcconnachie2009}, and all measurements show significant
structure on small scales indicative of a very rich merger history.
The BHB stars of the halo as measured with HST appear to paint the
same picture. The shape and mass of the stellar halo as measured with
BHB stars are broadly consistent with those of simulated halos that
contain a significant fraction of accreted stars. 

Support for this work was provided by NASA through grants GO-12055 and
through Hubble Fellowship grants 51273.01 awarded to K.M.G. from the
Space Telescope Science Institute, which is operated by the
Association of Universities for Research in Astronomy, Incorporated,
under NASA contract NAS5-26555.


\clearpage

\footnotesize
\begin{deluxetable}{lccccc}
\footnotesize
\tablecaption{Properties of the M31 Blue Horizontal Branch}
\tablehead{
\colhead{{\footnotesize Field}} &
\colhead{{\footnotesize R$_{med}$ (kpc)\tablenotemark{a}}} &
\colhead{{\footnotesize $F814W_{BHB}$\tablenotemark{b}}} &
\colhead{{\footnotesize $N_{BHB}$\tablenotemark{c}}} &
\colhead{{\footnotesize $N_{RGB}$\tablenotemark{d}}} &
\colhead{{\footnotesize $F_{BHB}$\tablenotemark{e}}}
}
\startdata
M31-B01-F07-WFC &  1.64 & \nodata & $<$2000 & 25300 & $\leq$0.08\\
M31-B01-F13-WFC &  1.70 &  \nodata & $<$2300 & 20000 & $\leq$0.12\\
M31-B01-F01-WFC &  1.70 &  \nodata & $<$3100 & 30300 & $\leq$0.10\\
M31-B02-F11-WFC &  2.47 &  \nodata & $<$3400 & 10500 & $\leq$0.32\\
M31-B02-F04-WFC &  2.96 &  \nodata & $<$3100 & 8500 & $\leq$0.36\\
M31-B02-F16-WFC &  2.95  &  25.233$\pm$0.018 &  1123$\pm$137 &  7240$\pm$85 & 0.16$\pm$0.01\\
M31-B02-F09-WFC &  3.38 &  25.240$\pm$0.024 & 878$\pm$139 &  6048$\pm$78 & 0.15$\pm$0.01\\
M31-B02-F15-WFC &  3.41 &  25.230$\pm$0.019 & 658$\pm$120 &  5276$\pm$73 & 0.12$\pm$0.01\\
M31-B02-F03-WFC &  3.42 &  25.196$\pm$0.024 & 929$\pm$143 &  6473$\pm$80 & 0.14$\pm$0.01\\
M31-B02-F08-WFC &  3.81 &  25.218$\pm$0.024  & 679$\pm$102 &  4279$\pm$65 & 0.16$\pm$0.01\\
M31-B02-F14-WFC &  3.84 &  25.245$\pm$0.021   & 738$\pm$88 &  3652$\pm$60 & 0.2$\pm$0.01\\
M31-B02-F02-WFC &  3.84 &  25.203$\pm$0.028 & 505$\pm$117 &  4865$\pm$70 & 0.1$\pm$0.01\\
M31-B02-F07-WFC &  4.25 & 25.257$\pm$0.017 & 633$\pm$67 &  2965$\pm$54 & 0.21$\pm$0.01\\
M31-B02-F01-WFC &  4.28 &  25.221$\pm$0.022 & 513$\pm$78 &  3395$\pm$58 & 0.15$\pm$0.01\\
M31-B02-F13-WFC &  4.28 &  25.238$\pm$0.018  & 603$\pm$65 &  2614$\pm$51 & 0.23$\pm$0.01\\
B08 halo11 & 11.0 & \nodata & 53$\pm$11 & 79$\pm$14 &  0.67$\pm$0.09\\
10394\_M31-HALO-NW & 17.7 & \nodata & 29$\pm$5 & 45$\pm$7 & 0.64$\pm$0.15 \\
B08 stream & 20.2 & \nodata & 11$\pm$3 & 22$\pm$5 &  0.50$\pm$0.10\\
B08 halo21 & 21.0 & \nodata & 14$\pm$4 & 20$\pm$4 &  0.70$\pm$0.24\\
11632\_M31-HALO-SE & 24.5 & \nodata & 8$\pm$3 & 15$\pm$4 & 0.53$\pm$0.23\\
B08 halo35ab & 35.0 & \nodata & 4$\pm$2 & 9$\pm$3 &  0.44$\pm$0.27\\ 
\enddata
\tablenotetext{a}{The median galactocentric distance of the stars in the the region kpc.}
\tablenotetext{b}{The F814W magnitude of the BHB at 0.1$<$F475W-F814W$<$0.5.}
\tablenotetext{c}{The number of BHB stars from 0.1$<$F475W-F814W$<$0.5 or equivalent.}
\tablenotetext{d}{The number of RGB stars from 1.5$<$F475W-F814W$<$3.5 and 22.0$<$F814W$<$22.5 or equivalent.}
\tablenotetext{e}{Fraction of BHB/RGB stars.}
\label{measurements}
\end{deluxetable}

\end{document}